\documentclass[11pt]{article}
\usepackage[margin=2.5cm]{geometry}
\usepackage[T1]{fontenc}
\usepackage{lmodern}
\usepackage{amsmath}
\usepackage{graphicx}
\usepackage{booktabs}
\usepackage{pgfplots}
\pgfplotsset{compat=1.18}
\usepackage[numbers,sort&compress]{natbib}
\usepackage[hidelinks]{hyperref}
\usepackage{xcolor}

\title{Classical visual illusions in vision-language models under an instruction ladder}
\author{Eiji Watanabe\\
\small Laboratory of Neurophysiology, National Institute for Basic Biology, Okazaki, Japan\\
\small \texttt{eijwat@gmail.com} \quad ORCID 0000-0001-9089-8122}
\date{}

\begin{document}
\maketitle

\begin{abstract}
Vision-language models (VLMs) often name the ``illusory'' answer when shown a classical visual illusion, but it is
unclear whether that answer comes from the picture or from knowledge about famous figures. We drew 300 new stimuli
of 25 classical illusions in three conditions---the illusion, the illusion with its inducing elements reversed, and a
control with the inducing elements removed---and asked about each image at eight rungs of an \emph{instruction ladder}
that runs from a false fact, through a neutral question and an instruction to report the physical truth, to telling
the model the name of the illusion and finally the true answer. Claude Sonnet 5.5 and Claude Haiku 5.5, with
thinking off and on, gave 460,800 answers. When asked for the physical fact, all four settings almost never gave the
illusion answer, and they resolved real 10\% differences in the same displays. When asked how the picture
\emph{looks}, the illusion answers returned, and they did not vanish after the true answer was stated. Two measures that cancel position preferences and recognition of the figure suggest that much of this depends on the stimulus: among side answers, Sonnet with thinking chose the illusion side in 85\% of cases after being told the
truth (95\% CI 74--96\%), with 16 of the 25 illusions showing a stimulus-driven effect. Yet for five figures---the
Fraser spiral, Kanizsa triangle, Pinna--Brelstaff rings, Ouchi pattern and, for Haiku, the rotating snakes---the
models gave the illusion answer to the control images as well, which suggests that they answered from the figure rather than from the effect. A false fact was mostly followed on control images but not on illusion images. The stimuli, prompts, all
responses and the code are released for testing further models; the Python code that draws the stimuli is
parametric and can generate new illusion images for other studies with models or human observers.
\end{abstract}

\section{Introduction}

Visual illusions are the classical probe of the assumptions built into a visual system: the Müller-Lyer lines look
unequal although they are equal, and the illusion persists after one has measured them. This persistence---the
illusion is not removed by knowing the truth---is what makes illusions informative about perception rather than
judgement \citep{pylyshyn1999,firestone2016}.

Deep networks reproduce several illusions \citep{watanabe2018,gomezvilla2019}, and vision-language models (VLMs) are
now tested on illusion images directly \citep{zhang2023grounding,shahgir2024illusionvqa,guan2024hallusionbench}.
A recurring difficulty is that the famous illusion figures are part of the training data, together with their
explanations. A model that answers ``the top line looks longer'' may do so because it sees a longer line, or because
it recognises the Müller-Lyer figure and knows what people report. VLMs even report illusions in images where the
illusion-inducing structure has been removed \citep{ullman2024illusion}.

Here we separate these possibilities with three design choices. (i) All stimuli are new drawings with varied colours,
sizes and orientations, and each illusion comes with a \emph{reversed} version, in which the inducing elements favour
the other answer, and a \emph{control} version without the inducing elements. (ii) The same image is asked about at
eight rungs of an instruction ladder, from a false fact to the stated truth, so that the effect of instructions and
knowledge can be traced. (iii) Measures are defined so that a preference for a position or a label, and recognition of
the figure's shape, do not count as an illusion. We report the responses of two current VLMs, each with thinking off
and on, and release everything needed to test more models.

We do not compare with human observers in this study. The ``illusion direction'' of every item is the direction
reported for that illusion in the perception literature, fixed when the stimulus was drawn.

\section{Methods}

\subsection{Stimuli}

\begin{figure}[t]
\centering
\includegraphics[width=\textwidth]{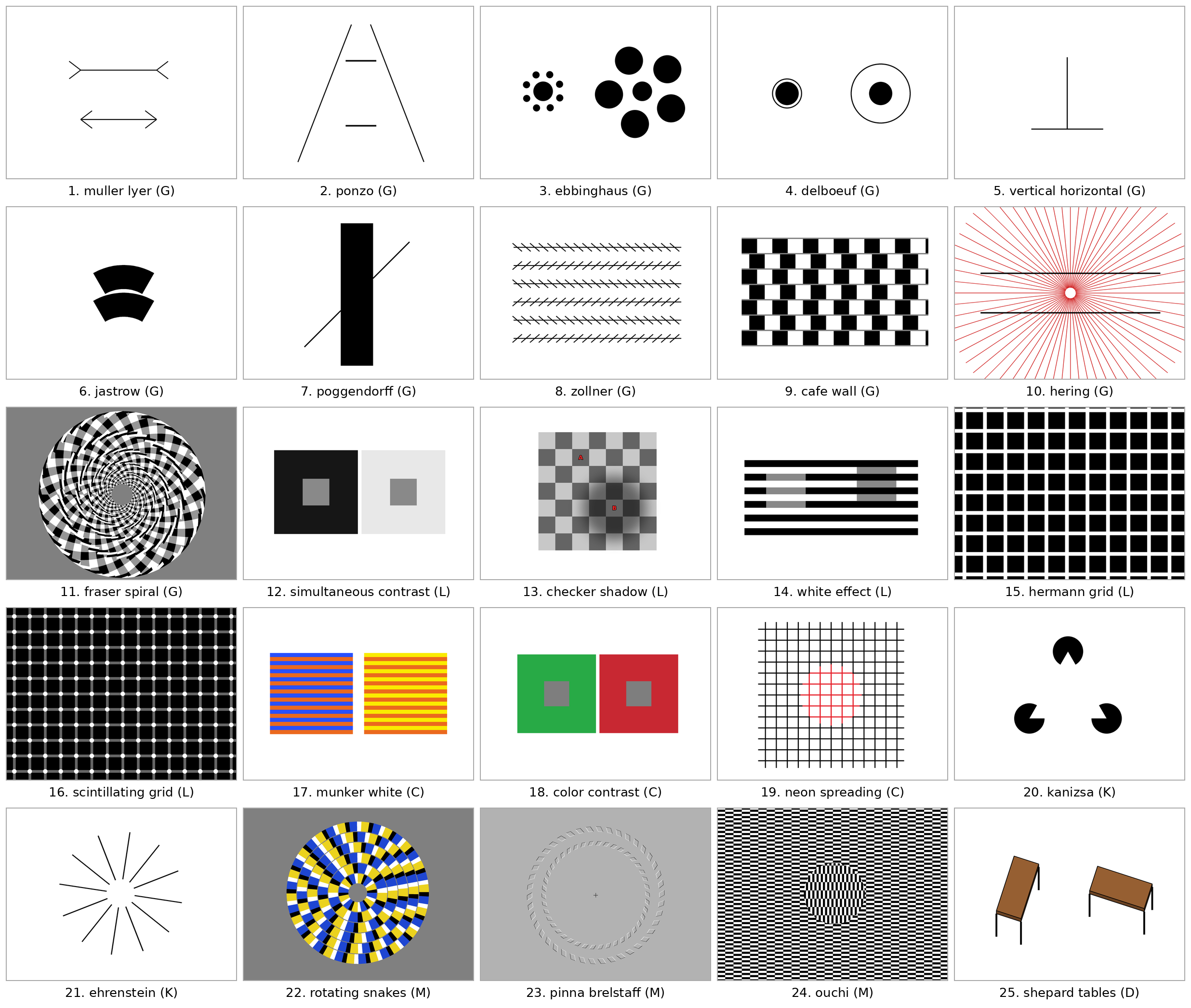}
\caption{The 25 illusions (normal condition, first render). Letters give the category: G geometric, L lightness,
C colour, K illusory contour, M motion, D depth.}
\label{fig:stimuli}
\end{figure}

Twenty-five illusions from six categories were drawn with Python and Pillow (Fig.~\ref{fig:stimuli}): geometric (G:
Müller-Lyer, Ponzo, Ebbinghaus, Delboeuf, vertical--horizontal, Jastrow, Poggendorff, Zöllner, café wall, Hering,
Fraser spiral), lightness (L: simultaneous contrast, checker shadow \citep{adelson1995}, White's effect
\citep{white1979}, Hermann grid, scintillating grid \citep{schrauf1997}), colour (C: Munker--White, colour contrast,
neon colour spreading \citep{vantuijl1975}), illusory contours (K: Kanizsa triangle \citep{kanizsa1955}, Ehrenstein),
motion in static images (M: rotating snakes \citep{kitaoka2003}, Pinna--Brelstaff rings \citep{pinna2000}, Ouchi
pattern \citep{ouchi1977}) and depth (D: Shepard tables \citep{shepard1990}). Figures with no single correct answer
(ambiguous figures, impossible objects, faces) were excluded.

\begin{figure}[t]
\centering
\includegraphics[width=0.75\textwidth]{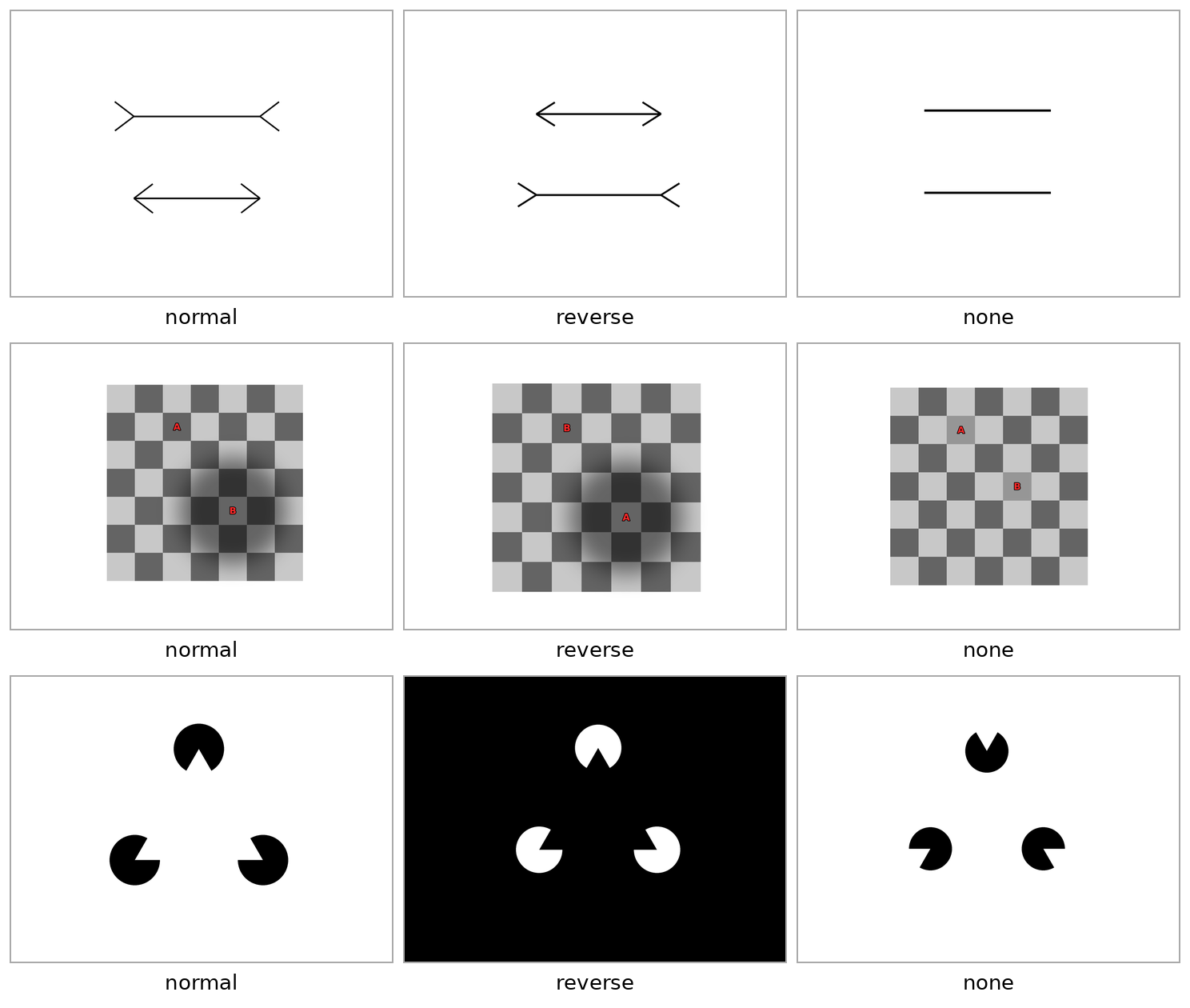}
\caption{The three conditions for three illusions. Rows: Müller-Lyer, checker shadow, Kanizsa triangle.}
\label{fig:conditions}
\end{figure}

Each illusion was drawn in three conditions (Fig.~\ref{fig:conditions}). \emph{Normal}: the illusion.
\emph{Reversed}: the inducing elements reversed. For the 13 two-sided items (which of two lines is longer, which grey
is lighter, \dots) this moves the illusion to the other side; for the 12 yes/no items (do the lines look parallel,
do you see a triangle, \dots) it is the other form of the illusion with the same expected answer (e.g.\ the Wundt form
of the Hering figure). \emph{None}: the inducing elements removed, so that the correct answer (``same'', ``no'',
``still'') is also what the picture looks like. Each condition was drawn in four renders that vary colours, line
width, size, position and (in half the renders) the axis of comparison, giving 300 images of $1200\times900$ pixels.
In 11 illusions with a measurable property, two of the four control renders contain a real 10\% difference (one line
longer, one disc larger, one grey lighter, \dots); these test whether the model can resolve the property at all.

\subsection{Instruction ladder}

Every image was asked about with the same question at eight rungs (example: Müller-Lyer, lines one above the other).
Rung $-1$, false fact: ``In fact, the bottom line is slightly longer than the other. Even so, which line looks longer?''
(the false side is always the one opposite the illusion; on control images it is chosen at random). Rung 0, neutral:
``Which line looks longer?'' Rung 1, warning: not to be fooled by appearances. Rung 2, physical fact: ``Which line is
actually longer?'' Rung 3: told that it is an illusion. Rung 4: told the kind of illusion (``an illusion of
length''). Rung 5: told its name. Rung 6, truth: ``The two lines are the same length. Even so, which line looks
longer?'' Except at rung 2, the question always asks how the picture \emph{looks}. Each prompt ends with the options
(``Answer with one word: top, bottom or same''). Each rung has three paraphrases that vary wording and politeness but
not the question. For yes/no items a false fact can only push towards the illusion (``the lines are not parallel''),
so their rung $-1$ is reported separately. All 921 distinct prompt texts are listed in the repository.

\subsection{Models and calls}

Claude Sonnet 5.5 (\texttt{claude-sonnet-5-5}) and Claude Haiku 5.5 (\texttt{claude-haiku-5-5}) were called through
the Anthropic Message Batches API in October 2026, each with thinking off and on. Thinking off: the lowest thinking
setting each model accepts, with effort \texttt{low}. Thinking on: adaptive thinking (the model decides whether and
how long to think) with effort \texttt{high}; these values were chosen in a pilot of 17,280 calls because lower
efforts produced almost no thinking. Every call was a fresh conversation with a one-sentence system prompt (``Answer
with exactly one word from the options given''), one image and one prompt, without tools, at temperature 1 and with
up to 8,000 output tokens. Each image $\times$ rung $\times$ paraphrase was asked 16 times: 300 images $\times$ 8
rungs $\times$ 3 paraphrases $\times$ 16 = 115,200 calls per setting, 460,800 in total. All calls succeeded;
answers were taken from the last word of the reply and were invalid in 481 calls (0.1\%), which count as neither
illusion nor correct answers. Haiku with thinking allowed thought in 99\% of the calls; Sonnet in 31\%.

\subsection{Measures}

For each setting, illusion and rung, $p_{\rm ill}$ is the share of illusion answers on the normal and reversed
images (the direction of the reversed images is the reversed one). $p_{\rm ill}$ alone is not a clean measure:
the wording ``\dots\ Even so, which looks \dots?'' at rungs $-1$ and $6$ invites an answer other than ``same'', and
a model may prefer a position (``top'') or a label. We therefore use two measures.

\emph{Two-sided items.} Among the answers that name a side, the share that names the illusion side,
$D = n_{\rm ill}/(n_{\rm ill}+n_{\rm other})$, pooled over normal and reversed images. Because the reversal moves the
illusion to the other side, a fixed preference for a side or label gives $D=0.5$; $D>0.5$ means the answers follow the
inducing elements.

\emph{Yes/no items.} The control images keep the overall shape of the figure, so a model that recognises the figure
may answer ``yes, it moves'' without any illusion in the picture. The stimulus effect is
$\Delta = p_{\rm ill}(\text{normal, reversed}) - p(\text{illusion word}\mid\text{control})$.

Confidence intervals are percentile bootstrap intervals (2,000 resamples) over image $\times$ paraphrase cells for
single illusions and over illusions for means across illusions. We call an illusion's effect \emph{stimulus-driven} at
a rung when the 95\% interval of $D$ lies above 0.5 (with at least 10\% side answers) or the interval of $\Delta$ lies
above 0.

\section{Results}

\subsection{Questions about the physical fact (rung 2)}

At rung 2 the share of illusion answers fell to 0.13 (Haiku, thinking off), 0.03 (Haiku, on), 0.004 (Sonnet, off)
and 0.006 (Sonnet, on) (Fig.~\ref{fig:curves}a). On the control images with a real 10\% difference the models chose
the physically larger side at rung 0 in 77\% (Haiku off), 100\% (Haiku on), 99\% (Sonnet off) and 100\% (Sonnet on) of
the calls. This suggests that the models can resolve the compared property in these displays and report it when asked for the physical fact.

\subsection{Questions about appearance (rungs 0, 1 and 3--6)}

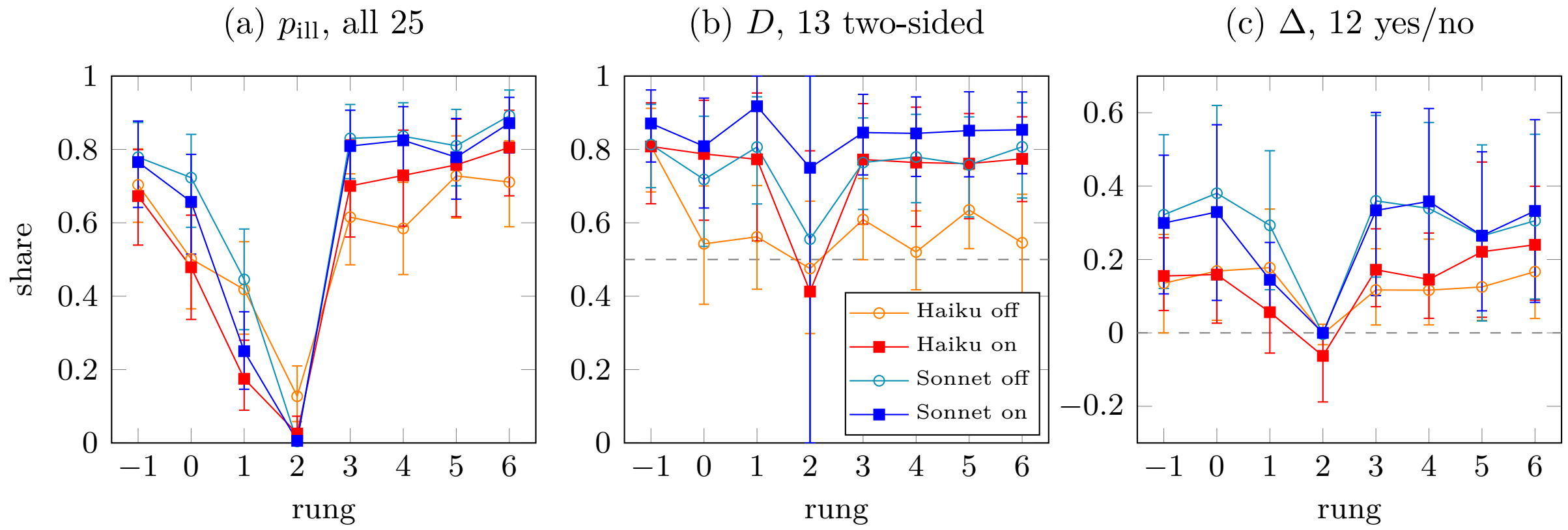
\begin{figure}[t]
\centering
\begin{tikzpicture}
\pgfplotsset{
  every axis/.style={width=0.355\textwidth, height=0.32\textwidth, xmin=-1.5, xmax=6.5, xtick={-1,0,...,6},
    xlabel={rung}, tick label style={font=\footnotesize}, label style={font=\footnotesize},
    title style={font=\small}},
  ser/.style={mark size=1.5pt, error bars/.cd, y dir=both, y explicit, error bar style={line width=0.4pt}},
}
\newcommand{\series}[4]{\addplot[#3, mark=#4, ser] table[x=rung, y=#2,
    y error plus expr=\thisrow{#2hi}-\thisrow{#2}, y error minus expr=\thisrow{#2}-\thisrow{#2lo}] {#1};}
\begin{axis}[name=a, title={(a) $p_{\rm ill}$, all 25}, ymin=0, ymax=1, ylabel={share}]
\series{curve_p.dat}{Haikuoff}{orange}{o}
\series{curve_p.dat}{Haikuon}{red}{square*}
\series{curve_p.dat}{Sonnetoff}{cyan!70!black}{o}
\series{curve_p.dat}{Sonneton}{blue}{square*}
\end{axis}
\begin{axis}[name=b, at={(a.east)}, anchor=west, xshift=0.9cm, title={(b) $D$, 13 two-sided}, ymin=0, ymax=1,
  legend style={font=\tiny, at={(0.98,0.02)}, anchor=south east, inner sep=1pt}, legend cell align=left]
\addplot[gray, dashed, domain=-1.5:6.5, forget plot] {0.5};
\series{curve_D.dat}{Haikuoff}{orange}{o}
\series{curve_D.dat}{Haikuon}{red}{square*}
\series{curve_D.dat}{Sonnetoff}{cyan!70!black}{o}
\series{curve_D.dat}{Sonneton}{blue}{square*}
\legend{Haiku off, Haiku on, Sonnet off, Sonnet on}
\end{axis}
\begin{axis}[name=c, at={(b.east)}, anchor=west, xshift=0.9cm, title={(c) $\Delta$, 12 yes/no}, ymin=-0.3, ymax=0.7]
\addplot[gray, dashed, domain=-1.5:6.5, forget plot] {0};
\series{curve_delta.dat}{Haikuoff}{orange}{o}
\series{curve_delta.dat}{Haikuon}{red}{square*}
\series{curve_delta.dat}{Sonnetoff}{cyan!70!black}{o}
\series{curve_delta.dat}{Sonneton}{blue}{square*}
\end{axis}
\end{tikzpicture}
\caption{The instruction ladder. Rung $-1$ false fact, 0 neutral, 1 warning, 2 physical fact, 3 illusion, 4 kind,
5 name, 6 truth. (a) Share of illusion answers. (b) Direction among side answers (0.5 = no direction; at rung 2 few
side answers remain, hence the wide intervals). (c) Stimulus effect for yes/no items. Means over illusions with 95\%
bootstrap intervals over illusions.}
\label{fig:curves}
\end{figure}

As soon as the question asked about appearance, illusion answers returned (Fig.~\ref{fig:curves}a). They were not
reduced by telling the model that the figure is an illusion (rung 3), its kind (rung 4) or its name (rung 5), nor by
stating the true answer (rung 6): $p_{\rm ill}$ at rung 6 was 0.71, 0.81, 0.89 and 0.87 for the four settings. Rung 1
(a warning not to be fooled) was the only appearance question that reduced illusion answers, most strongly with
thinking.

The direction measure $D$ suggests that this is not only a side preference (Fig.~\ref{fig:curves}b). For Sonnet with thinking
$D$ was 0.81 [0.64, 0.94] at rung 0 and 0.85 [0.74, 0.96] at rung 6; for Haiku with thinking 0.79 and 0.77. Haiku
without thinking was near chance (0.54 and 0.55): its side answers did not follow the inducing elements. For the
yes/no items the stimulus effect $\Delta$ was smaller, 0.33 [0.08, 0.57] for Sonnet with thinking at rung 6
(Fig.~\ref{fig:curves}c).

Counting illusions with a stimulus-driven effect at rung 6 gives 16 of 25 for Sonnet with thinking, 13 for Sonnet
without, 14 for Haiku with and 6 for Haiku without thinking (Table~\ref{tab:illusions}). Effects that were stimulus-driven in all three settings other than Haiku without thinking were found for the Müller-Lyer, Ebbinghaus, Delboeuf, simultaneous contrast, checker shadow, colour contrast
and Shepard tables (two-sided), and the Zöllner, Hering, Hermann grid, Munker--White and Ehrenstein figures (yes/no).
The vertical--horizontal and Jastrow figures produced no stimulus-driven direction at rung 6 in any setting, and the Ponzo figure only for Haiku with thinking. Values of $D$ well below 0.5 mean answers consistently \emph{against} the illusion; this occurred for Haiku without
thinking (colour contrast, Delboeuf) and for Haiku with thinking on the Poggendorff figure.

\subsection{Answers on the control images}

For five yes/no figures the control images received the illusion answer as often as the illusions themselves
(Table~\ref{tab:illusions}, control column): the Fraser spiral without twist was called a spiral, the Pinna--Brelstaff
rings without luminance polarity and the Ouchi pattern without the orientation difference were said to move or
shimmer (94--100\% of calls in every setting), and the Kanizsa figure with outward-facing inducers was said to contain a triangle (90--100\%). In a check before the main run, the models' thinking summaries named the illusion and its ``typical'' answer before answering. For the Kanizsa figure Sonnet reported a triangle on the control images even more often than on the
illusion images ($\Delta<0$). The rotating-snakes control (alternating luminance order, which humans do not see
moving) was said to rotate in 92--100\% of calls, but here the two-sided measure separates knowledge from the
stimulus: Sonnet with thinking named the rotation direction predicted by the luminance order of the normal and reversed images ($D=0.99$ at rung 0; 0.61 without thinking), whereas Haiku chose one direction regardless ($D=0.50$--$0.53$).

\subsection{The false fact (rung $-1$)}

At rung $-1$ the models were told that the side opposite the illusion is slightly larger and then asked which looks
larger. On the illusion images, the side answers went to the illusion side, against the false fact ($D=0.81$--$0.87$).
The same wording on the control images, where the false side is random, gives an indication of how the models treat the statement itself: Sonnet named the stated side (share of the other side 0.16 [0.04, 0.32] without and 0.16 [0.03, 0.34] with
thinking), whereas Haiku without thinking tended to name the other side (0.69), apparently reading ``even so'' as a
contrast. For Sonnet, this suggests that the illusion images can override a statement that the model otherwise tends to follow.

\subsection{Thinking off and on}

Allowing thinking strengthened the direction of side answers (Fig.~\ref{fig:curves}b) and made the warning at rung 1
more effective (Fig.~\ref{fig:curves}a). For Sonnet, which thought in only 31\% of calls when allowed to, the calls
with thinking had a higher $D$ than those without at the neutral rung (0.93 vs 0.88) and at the warning rung (0.93 vs
0.86), and similar values at rung 6 (0.87 vs 0.85). Thinking did not remove illusion answers once the question asked
about appearance.

\begin{table}[p]
\centering
\scriptsize
\setlength{\tabcolsep}{3pt}
\begin{tabular}{llccccccccccc}
\toprule
 & & & \multicolumn{2}{c}{Haiku/off} & \multicolumn{2}{c}{Haiku/on} & \multicolumn{2}{c}{Sonnet/off} & \multicolumn{2}{c}{Sonnet/on} & control & $\pm$10\% \\
\cmidrule(lr){4-5} \cmidrule(lr){6-7} \cmidrule(lr){8-9} \cmidrule(lr){10-11}
Illusion & cat. & meas. & r0 & r6 & r0 & r6 & r0 & r6 & r0 & r6 & r0 & r0 \\
\midrule
M\"uller-Lyer & G & $D$ & 0.47 & 0.41 & 0.84 & 0.81 & 1.00 & 1.00 & 1.00 & 1.00 & 1.00 & 1.00 \\
Ponzo & G & $D$ & 0.74$^\dagger$ & 0.52 & 0.88$^\dagger$ & 0.75 & 0.50 & 0.50 & 0.50 & 0.46 & 1.00 & 0.99 \\
Ebbinghaus & G & $D$ & 0.48 & 0.64 & 1.00 & 0.93 & 1.00 & 1.00 & 1.00 & 1.00 & 1.00 & 1.00 \\
Delboeuf & G & $D$ & 0.05 & 0.22 & 0.97 & 1.00 & 0.89 & 1.00 & 0.96 & 1.00 & 1.00 & 0.99 \\
Vertical--horizontal & G & $D$ & 0.26 & 0.63 & 0.76$^\dagger$ & 0.50 & 0.59 & 0.50 & 0.36 & 0.50 & 0.57 & 0.87 \\
Jastrow & G & $D$ & 0.51 & 0.46 & -- & 0.54 & 0.23 & 0.38 & 0.13$^\dagger$ & 0.50 & 1.00 & 1.00 \\
Poggendorff & G & $D$ & 0.87 & 0.98 & 0.00 & 0.41 & 0.00$^\dagger$ & 0.58 & 0.83$^\dagger$ & 0.78 & 0.99 & 0.97 \\
Simultaneous contrast & L & $D$ & 0.94 & 0.79 & 1.00 & 0.97 & 1.00 & 1.00 & 1.00 & 1.00 & 1.00 & 0.96 \\
Checker shadow & L & $D$ & 0.48 & 0.53 & 0.93 & 1.00 & 0.99 & 1.00 & 1.00 & 1.00 & 0.86 & 1.00 \\
White's effect & L & $D$ & 0.79 & 0.42 & 0.55 & 0.65 & 0.53 & 0.59 & 0.83 & 0.88 & 0.95 & 0.82 \\
Colour contrast & C & $D$ & 0.00 & 0.01 & 1.00 & 1.00 & 1.00 & 1.00 & 1.00 & 1.00 & 1.00 & -- \\
Rotating snakes & M & $D$ & 0.50 & 0.50 & 0.53 & 0.51 & 0.61 & 0.94 & 0.99 & 0.98 & 0.02 & -- \\
Shepard tables & D & $D$ & 0.97 & 0.99 & 1.00 & 1.00 & 1.00 & 1.00 & 0.91 & 0.99 & 0.98 & 0.73 \\
\midrule
Z\"ollner & G & $\Delta$ & $-$0.01 & $-$0.03 & 0.10 & 0.58 & 0.76 & 0.34 & 0.83 & 0.37 & 0.13 & -- \\
Caf\'e wall & G & $\Delta$ & 0.89 & 0.14 & 0.49 & 0.02 & 0.30 & 0.14 & 0.38 & 0.47 & 0.28 & -- \\
Hering & G & $\Delta$ & 0.46 & 0.03 & 0.49 & 0.61 & 1.00 & 0.74 & 0.93 & 0.89 & 0.00 & -- \\
Fraser spiral & G & $\Delta$ & 0.05 & 0.00 & 0.01 & 0.00 & 0.00 & 0.00 & 0.00 & 0.00 & 0.99 & -- \\
Hermann grid & L & $\Delta$ & 0.28 & 0.73 & 0.78 & 0.21 & 0.82 & 0.56 & 0.53 & 0.70 & 0.13 & -- \\
Scintillating grid & L & $\Delta$ & 0.07 & 0.02 & 0.00 & 0.03 & 0.01 & $-$0.03 & 0.00 & $-$0.28 & 0.00 & -- \\
Munker--White & C & $\Delta$ & 0.00 & 0.38 & 0.13 & 0.73 & 0.72 & 0.99 & 0.79 & 1.00 & 0.00 & -- \\
Neon spreading & C & $\Delta$ & 0.33 & 0.15 & 0.02 & 0.18 & 0.46 & 0.00 & 0.14 & 0.13 & 0.19 & -- \\
Kanizsa triangle & K & $\Delta$ & 0.03 & 0.00 & $-$0.08 & 0.02 & $-$0.37 & $-$0.07 & $-$0.45 & $-$0.21 & 0.94 & -- \\
Ehrenstein & K & $\Delta$ & $-$0.13 & 0.58 & $-$0.03 & 0.50 & 0.88 & 0.99 & 0.80 & 0.92 & 0.45 & -- \\
Pinna--Brelstaff & M & $\Delta$ & 0.06 & 0.00 & 0.00 & 0.00 & 0.00 & 0.00 & 0.00 & 0.00 & 0.98 & -- \\
Ouchi & M & $\Delta$ & 0.01 & 0.00 & 0.00 & 0.00 & 0.00 & 0.00 & 0.00 & 0.00 & 1.00 & -- \\
\bottomrule
\end{tabular}

\caption{Per illusion: the direction $D$ (two-sided items) or stimulus effect $\Delta$ (yes/no items) at rung 0
(neutral) and rung 6 (truth) for the four settings. $^\dagger$: fewer than 10\% side answers, so $D$ rests on few
answers; --: no side answers. Control, r0: for two-sided items the share of the correct answer (``same'') on control
images, for yes/no items the share of the illusion word on control images (high values mean the figure is answered
from its shape), mean of the four settings. $\pm$10\%: share of correct answers on the control images with a real
10\% difference, mean of the four settings.}
\label{tab:illusions}
\end{table}

\section{Discussion}

Two current VLMs reported many classical illusions in the direction known from human perception, and---as for human
observers---these answers survived being told the truth. Three observations suggest that this may not be only a matter of remembered facts about famous figures. The stimuli were new drawings with varied layout; the answers followed the
inducing elements when they were reversed, which cancels any preference for a side or label; and the same displays
with a real 10\% difference, or the question about the physical fact, were answered correctly, which makes it unlikely that the models simply could not see the difference.

At the same time, recognition of a figure appears to be a second route to the illusion answer. For figures whose
illusion is motion or an illusory surface (Fraser spiral, Pinna--Brelstaff, Ouchi, Kanizsa) the control images, which
do not induce the illusion in humans, received the same answers, consistent with reports of VLMs seeing illusions
where there are none \citep{ullman2024illusion}. Measures without such controls might count these as illusions.

The instruction ladder also helps to separate the effects of instruction. Only the question about the physical fact (rung 2)
removed illusion answers; information about the illusion (rungs 3--5) did not, and the true answer (rung 6) did not.
This dissociation between ``what is'' and ``what looks'' resembles the cognitive impenetrability of human illusions
\citep{pylyshyn1999,firestone2016}, but it may equally reflect that the models have learnt how people describe the
appearance of such figures. Our data cannot tell these apart. As VLMs are increasingly applied to many kinds of image analysis, studying such human-like biases is likely to become increasingly important.

\paragraph{Limitations.} (1) No human data were collected; the illusion direction comes from the literature and some
of our renders may induce weaker or different effects in people (for the reversed vertical--horizontal figure the
predicted direction is uncertain). (2) Two models from one developer were tested. (3) The phrasing ``Even so, which
looks \dots?'' at rungs $-1$ and 6 invites a side answer; we addressed this with $D$ and the control images, but other
phrasings could change $p_{\rm ill}$. (4) Answers were restricted to one word, and the thinking of the models is
available only as summaries. (5) Temperature 1 and 16 repetitions estimate the answer distribution of each prompt, not
the variation between model versions.

\section*{Data and code availability}
Stimuli, prompts, all 478,880 responses (main run, pilot and pre-run check), analysis code and results are available
at \url{https://github.com/eijwat/vlm-illusion-ladder} and archived at Zenodo (\href{https://doi.org/10.5281/zenodo.24390082}{doi:10.5281/zenodo.24390082}). The stimulus code
(\texttt{ladder/stimuli.py}, Python and Pillow only) draws each of the 25 illusions, together with its reversed and control versions,
from a few drawing parameters (size, line width, colours, orientation, and illusion-specific values such as the fin
length and angle of the Müller-Lyer figure) and a fixed random seed. Changing these values or the seed gives new
renders, so the code can be reused to make new stimulus sets, for example for psychophysics with human observers or for
training and testing other models. Code: MIT licence;
stimuli, prompts, data and results: CC BY 4.0.

\section*{Acknowledgements}
This work was supported by the Ministry of Education, Culture, Sports, Science and Technology (MEXT), Japan, through
SPReAD (Supporting Pioneering Research through AI for 1,000 Discovery challenges),.

The stimulus and analysis code, the analysis and drafts of this manuscript were prepared with the help of Claude
Code (Anthropic), an AI coding assistant, under the author's direction. The author designed the study, checked the
stimuli and is responsible for the content.

\bibliographystyle{unsrtnat}
\bibliography{refs}

\end{document}